\documentstyle[12pt,epsfig,psfig]{elsart}

\begin{document}
\begin{frontmatter}
\title{The STAR  Photon Multiplicity Detector}


\author[panjab]{M. M. Aggarwal},
\author[jammu]{S. K. Badyal},
\author[vec]{P. Bhaskar}, 
\author[panjab]{V. S. Bhatia},
\author[vec]{S. Chattopadhyay},
\author[vec]{S. Das},
\author[jammu]{R. Datta},
\author[iop]{A. K. Dubey},
\author[vec]{M. R. Dutta~Majumdar},
\author[vec]{M. S. Ganti},
\author[vec]{P. Ghosh},
\author[jammu]{A. Gupta},
\author[jammu]{M. Gupta},
\author[jammu]{R. Gupta},
\author[panjab]{I. Kaur},
\author[panjab]{A. Kumar},
\author[jammu]{S. Mahajan},
\author[iop]{D. P. Mahapatra},
\author[jammu]{L. K. Mangotra},
\author[iop]{D. Mishra},
\author[iop]{B. Mohanty},
\author[jammu]{S. K. Nayak},
\author[vec]{T. K. Nayak},
\author[vec]{S. K. Pal},
\author[iop]{S. C. Phatak},
\author[jammu]{B. V. K. S. Potukuchi},
\author[jai]{R. Raniwala},
\author[jai]{S. Raniwala},
\author[iop]{R. Sahoo},
\author[jammu]{A. Sharma},
\author[vec]{R. N. Singaraju},
\author[panjab]{G. Sood},
\author[vec]{M. D. Trivedi},
\author[iit]{R. Varma},
\author[vec]{Y. P. Viyogi}

\address[panjab]{Physics Department, Panjab University, Chandigarh 160014, India}
 \address[jammu]{Physics Department, Jammu University, Jammu 180001, India}
 \address[vec]{Variable Energy Cyclotron Centre, Kolkata 700 064, India}
 \address[iop]{Institute of Physics, Bhubaneswar 751005, India}
 \address[jai]{Physics Department, Rajasthan University, Jaipur 302004, India}
\address[iit]{Indian Institute of Technology, Mumbai 400076, India}

\begin{abstract}
  
  Details concerning  the design, fabrication and  performance of STAR
  Photon Multiplicity Detector (PMD) are presented. The PMD will cover
  the forward region, within the pseudorapidity range 2.3--3.5, behind
  the  forward time projection  chamber. It  will measure  the spatial
  distribution  of   photons  in  order  to   study  collective  flow,
  fluctuation and chiral symmetry restoration.

\end{abstract}
\end{frontmatter}

\section{Introduction}

\begin{figure}
\begin{center}
\epsfig{figure=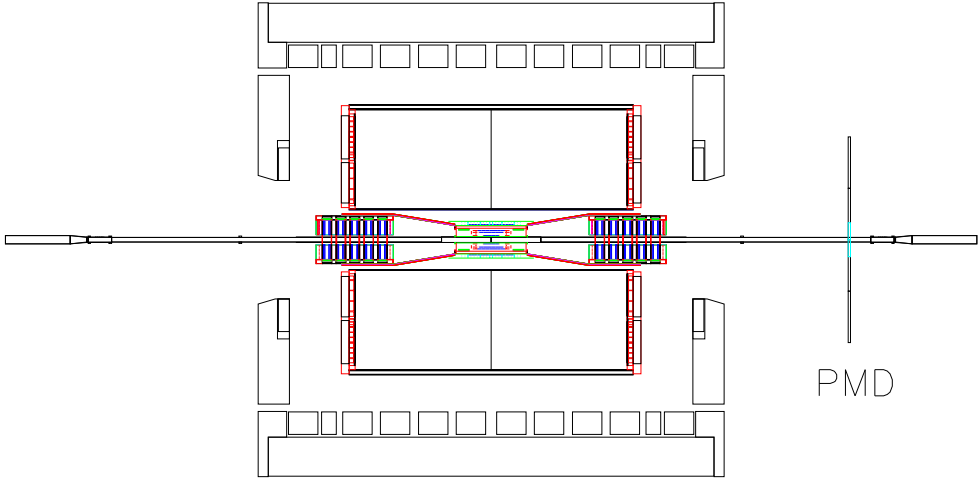,width=13cm}
\caption{The PMD in the STAR set up relative to central detector
TPC. The PMD is located at 550~cm from vertex and kept outside
the STAR magnet.}
\end{center}
\label{starpmd}
\end{figure}

A preshower Photon Multiplicity Detector (PMD) will be installed on
the east wall of the wide angle hall in 2002 shutdown period in STAR.
This detector is designed to measure photon multiplicity in the
forward region where high particle density precludes the use of a
calorimeter.  Fig~\ref{starpmd} shows the PMD relative to other
detectors within the STAR setup as implemented through GEANT
simulation.  The inclusion of the PMD enhances the phase space
coverage of STAR with photons considerably, in pseudorapidity ($\eta$
= -- ln~tan$\theta$/2) upto $\eta$ = 3.5 with full azimuthal
acceptance and in $p_T$ down to about 25 MeV/c
\cite{starnote,startdr}.

Using the measurement of multiplicity and spatial distribution of
photons on an event by event basis and combining the information from
other detectors, the PMD will be able to address the following broad
topics in physics : (a) determination of reaction plane and the probes
of thermalisation via studies of azimuthal anisotropy and flow, (b)
critical phenomena near the phase boundary leading to fluctuations in
global observables like multiplicity and pseudorapidity distributions,
and (c) signals of chiral symmetry restoration (e.g., disoriented
chiral condensates).
 
The basic principle of the measurement of photon multiplicity using
the PMD is similar to those of preshower detectors used in WA93 and
WA98 experiments at CERN SPS \cite{wa93pmd,wa98pmd}.  It consists of
highly segmented detector placed behind a lead converter of suitable
thickness. A photon produces an electromagnetic shower on passing
through the converter. These shower particles produce signals in
several cells of the sensitive volume of the detector. Charged hadrons
usually affect only one cell and produce a signal resembling those of
Minimum Ionizing Particles (MIPs). The thickness of the converter is
optimized such that the conversion probability of photons is high and
transverse shower spread is small to minimize shower overlap in a high
multiplicity environment.  In order to have better hadron rejection
capability, another plane of the detector of identical dimension as of
the preshower part is placed before the lead plate, which acts as a
veto for charged particles.

The detector is based on a proportional counter design using Ar +
CO$_2$ gas mixture. This gas mixture is preferred because of its
insensitivity to neutrons.  To handle the high particle density in the
forward region, the detector technology has been chosen with the
considerations that (i) multihit probability should be less (ii) MIP
should be contained in one cell, (iii) low energy $\delta$-electrons
should be prevented from travelling to nearby cells and causing
cross-talk among adjacent cells.  Requirement of granularity and
isolation of cells require the segmentation of the detector gas volume
with material effective for reducing $\delta$-electrons from crossing
one cell to other. We have used honeycomb cellular geometry with wire
readout. The copper honeycomb body forms the common cathode and is
kept at a large negative potential. It also supports the printed
circuit boards (PCBs) which are used for signal collection and for
extension of the cathode required for proper field shaping. Details
can be found in \cite{alicetdr,startdr}.

The present article is organised as follows.  The detector hardware
and support structure are described in Section 2. The front-end
electronics and readout scheme are described in Section 3.
Performance of the prototypes and a comparison with simulation results
are described in Section 4. A summary is presented in Section 5.

\section{The Detector}

The detector consists of an array of hexagonal cells.  A unit cell is
shown schematically in Fig.~ 2(a) along with a longitudinal section
illustrating the use of extended cathode for field shaping. This
design was arrived at after several simulation studies and prototype
tests and ensures uniform charged particle detection efficiency
throughout the cell \cite{alicenim}.

A honeycomb of 24$\times$24 cells forms a unit module. This is a
rhombus of side approx. 254~mm having identical boundaries on all the
four sides. Cell walls at the boundary are kept half as thick as those
inside so that adjacent unit modules join seamlessly.

A set of unit modules are enclosed in a gas-tight chamber called
supermodules.  The number of unit modules varies from 4 to 9 within a
supermodule.  The STAR PMD consists of 24 supermodules arranged in the
form of a hexagon as shown in Fig.~2(b). This geometry ensures full
azimuthal coverage with minimum number of supermodules.

\begin{figure}
\begin{center}
\epsfig{figure=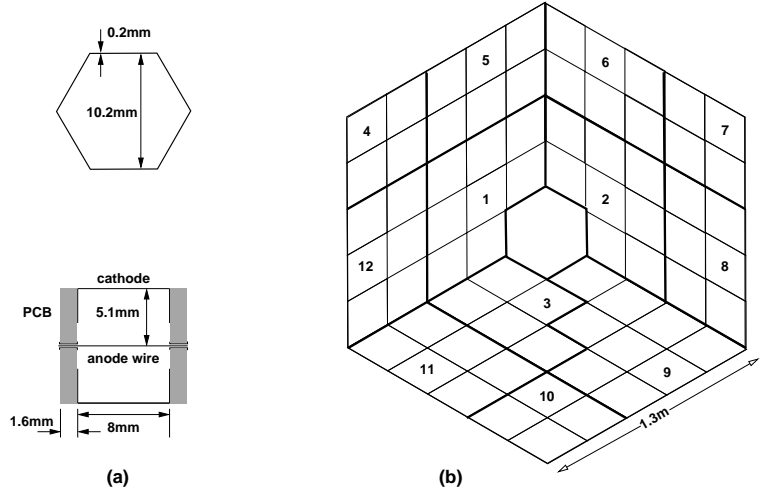,height=6.5cm,width=10cm}
\caption{ (a) Unit cell schematic with cross-section showing the
  dimensions and the cathode extension, (b) Layout of the STAR PMD.
  Thick lines indicate supermodule boundaries. There are 12
  supermodules each in the preshower plane and the veto plane.
  Divisions within a supermodule denote unit modules.}
\end{center}
\label{detector}
\end{figure}

\subsection{Unit Module}

\begin{figure}
\begin{center}
\epsfig{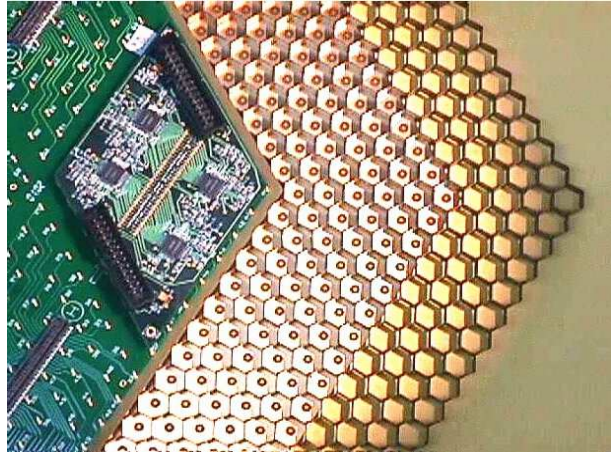}
\caption{ Components of a unit module : Copper honeycomb, placed 
  between two PCBs. The top PCB is seen with connectors and a FEE
  board. The cathode extension on the inside of the bottom PCB and the
  island separating the anode wire with the cathode is visible through
  the honeycomb. The photograph was taken with unassembled
  components.}
\end{center}
\label{unit-module}
\end{figure}

The components of a unit module are shown in Fig.~3.  It consists of a
custom-built copper honeycomb sandwiched between two PCBs which hold
the anode wire and provide extension to cathode.  The top PCB,
containing the electronics boards, has solder-islands at the centre of
each cell with a 0.5~mm dia gold-plated through-hole. Signal tracks
from cluster of 64 cells are brought to a 70-pin connector. The PCBs
on the bottom side have only soldering islands without signal tracks,
serving as anchor points. The inner part of the PCBs are gold-plated,
with circular islands near the anode wire and form part of the
extended cathode.

A copper unit cell is the building block of the honeycomb. It is
fabricated using 0.2 mm thick ETP grade copper sheets which are
solder-coated on one side. The sheet is cut to precise dimensions
along with notches and bent in hexagonal form with precision dies.
These are arranged in a 24$\times$24 matrix in a high precision jig of
rhombus shape. Hexagonal Stainless Steel inserts, having dimensions
matching the inner dimensions of the cell, are inserted in each cell.
The assembly is heated so that soldered surfaces join to form a rigid
honeycomb.

The honeycomb, after cleaning, is dip-coated with high conductivity
graphite paint having thickness of $\sim$10~$\mu$m.  The unit
honeycomb module has stiff 1~mm dia. brass screws situated at 24
different locations, which act as guides for attaching the PCBs on
both sides, ensuring proper alignment.  They are also used to bring
out the high voltage connections of the cathode onto the PCBs.  The
two PCBs are attached on both sides of the honeycomb, aligning with
the screws. These screws protrude only 0.5~mm above the PCB surface
and are fixed with thin nutss on the surrounding islands.  The islands
are covered with ABS plastic caps.

The gold-plated tungsten wires (20 $\mu$m dia.) are inserted through
the holes on the PCBs, using a needle and a tensioning jig.  After
applying tension of $\sim$ 30\% of the elastic limit, the wires are
soldered onto the islands on the PCBs about 3~mm away from the hole
(for details see Ref.~\cite{alicetdr}).  The plated through-holes,
where wires emerge, are then closed with a tiny amount of fast-setting
epoxy to make them gas-tight. This scheme prevents creepage of solder
flux into the cell and makes soldering easier.

A moulded FR4 edge frame is bonded to the top PCB.  This frame has a
beveled outer wall which forms a V-shaped groove between the
boundaries of the adjoining unit modules.

Quality assessment for the fabrication of the unit module is done by
several ways, viz, visual inspection of the solder joints and epoxy
filling in the holes and measurement of resistance of each wire to
monitor dry-soldering contacts. Resistance measurement shows that the
RMS is within 5$\%$ for one unit module.
In addition, high voltage tests are also performed after connecting
the front-end electronics boards and the pedestals of chips monitored
to test stable operation of the detector.


\subsection{Supermodule}

Supermodule is a gas-tight chamber made of 3 mm thick FR4 grade glass
epoxy sheet as the base plate and a 7 mm thick and 25 mm high aluminum
boundary wall.  A schematic cross-section of a supermodule is shown in
Fig.~4.  The opposite sides of the boundary walls have gas-feed
channels. Each channel has 24 openings into the chamber.  This scheme,
along with the notches in the cells, keep the gas flow impedance low.
A set of assembled unit modules are placed to fill the inner area of
the supermodule enclosure, leaving a gap of 1 mm on all sides to
accommodate general assembly tolerance and to provide insulation
between the honeycomb cathode and the boundary. Teflon spacers are
inserted into this gap all along the boundary to arrest any movement
of the unit modules and also to insulate the honeycomb cathode from
the walls.  The groove formed at the junctions of all the unit modules
and between the boundary walls and the unit modules are filled with
high viscosity silicone adhesive compound to make the chamber
gas-tight.

Gas is fed through the connector at the end of the long gas feed
channel. It enters through all the entry points in the channel
simultaneously, at the depth of 4 mm from the bottom of the chamber.
It then flows through the notches and exits at the other edge of the
supermodule through the 24 openings of the output channel.  An
aluminum enclosure containing one SHV connector, an HV limiting
resistor and decoupling capacitor is now fixed at one corner of the
supermodule very close to the HV tapping point.

\begin{figure}
\begin{center}
\epsfig{figure=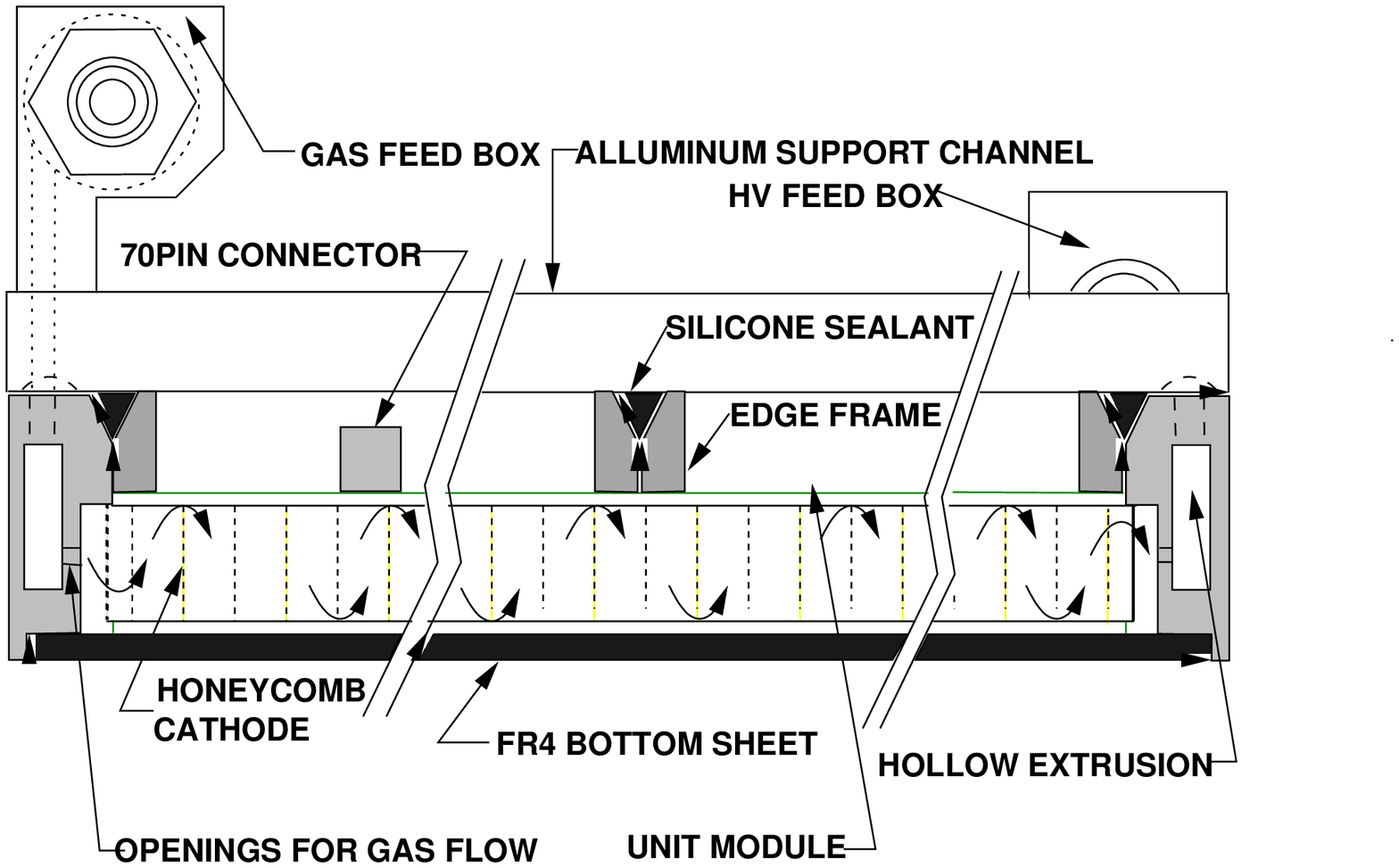,height=6.75cm,width=10.2cm}
\caption{Schematic cross-section of a supermodule showing the boundary walls,
  gas flow channels, high voltage connection and gas-tight sealings.}
\end{center}
\label{smsection}
\end{figure}

\subsection{Support Structure}

The drawing of the support structure is shown in Fig.~5.  It has two
parts: (a) the support plates, and (b) the suspension movement
mechanisms.  A 5~mm thick flat stainless steel plate is used to
support the lead converter plates and supermodules in each half of the
PMD.  It has tapped holes for screws corresponding to hole positions
in the lead plates and in the supermodules.  The lead converter plates
are sandwiched between two layers of gas detectors.

\begin{figure}
\begin{center}
\epsfig{figure=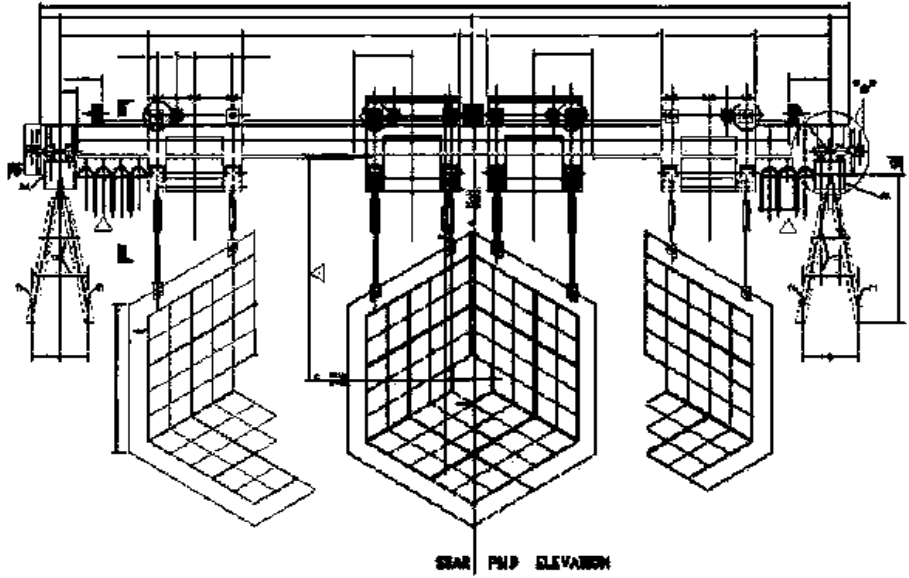,height=7.75cm,width=12.0cm}
\caption{PMD support mechanism. The inner hexagonal part shows the two halves 
joined during data taking operation. The two halves, when separated for 
servicing, look as shown on the right and left. } 
 
\end{center}
\label{support}
\end{figure}

The two halves of the detector are supported on the girders and hang
freely in a vertical position.  The support structure allows both $x$-
and $z$- movements of the detector.  Each half of the detector can be
separated for access by a smooth independent movement controlled by
limit switches.  The hanging elements have free swinging pivots, fine
adjustments for horizontal motion, and plane position adjustments for
alignment of the detector. The services of the two halves are also
independent.  When fully open, the two halves provide sufficient
clearance for the poletip support of the STAR magnet to move in.

The edges of the support plate are also used for mounting the gas feed
manifolds, show boxes for low voltages supplies and general support
for distribution of cables onto the detector.

\section{The Front End Electronics and Readout }

The front-end electronics for processing the PMD signals is based on
the use of 16-channel GASSIPLEX chips developed at CERN
\cite{gassiplex} which provide analog multiplexed signals and readout
using the custom built ADC board (C-RAMS)\footnote{Obtained from
    CAEN, Italy}. C-RAMS can handle a maximum of 2000 multiplexed
signals.  Considering the symmetry requirements of the detector
hardware, the readout of the entire PMD has been divided into 48
chains. Each chain covers three unit modules and has 1728 channels.
  
Each readout chain is driven by (i) a translator board (ii) 27 FEE
boards each consisting of 4 GASSIPLEX chips and (iii) a buffer
amplifier board.
  
(i) Translator Board: It converts NIM levels of all control signals
into the level required for the operation of GASSIPLEX chips.
Operating voltage for these chips is $\pm$2.75V and hence all the NIM
signals are to be translated to 0 to 2.75 V levels.
  
(ii) FEE board: The cells in the unit modules are arranged in clusters
consisting of 8 $\times$ 8 cells connected to a 70-pin connector.
This cluster of 64 cells is read out by a FEE having four GASSIPLEX
chips.  One such board is shown in Fig.~6. For geometrical
considerations the FEE board is also made in rhombus shape.  When all
the boards are placed on the detector, they almost fully cover the
detector area. This arrangement helps to reduce the material and also
provides a ground shield for the detector.
 
To reduce voltage drops over a long chain of 1728 channels, a bus-bar
like design has been adopted to provide power to the FEE boards.  To
protect the input channels against high voltage spikes, a provision
has been made on the board layout to connect a diode protection
circuit.

\begin{figure}[htbp]
\begin{center}
\epsfig{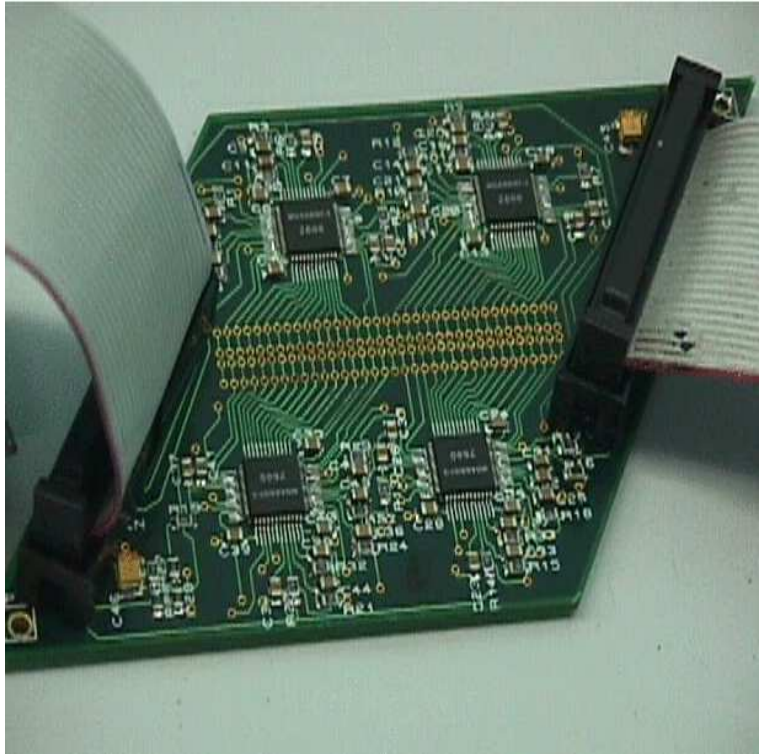}
\caption{ Photograph of a FEE board with four GASSIPLEX chips.}
\end{center}
\label{4chip}
\end{figure}

(iii) Buffer amplifier board: The buffer amplifier is used for the
transmission of a train of analog multiplexed signals to the readout
module via a low impedance cable.

Digitization using C-RAMS requires that all multiplexed pulses within
a chain should have the same polarity.  In order to read the full
chain, the pedestals in the chain need to be adjusted to the minimum
of the pedestals in the chain. This shifting of the pedestal
effectively reduces the dynamic range.  To minimize the reduction in
dynamic range due to pedestal adjustment, we need to select the chips
for a chain having minimum pedestals in very close range.

For proper quality control in the assembly of FEE boards, each
GASSIPLEX chip has been tested for full functionality of each channel.
In addition the pedestals of all the channels have been measured. The
minimum pedestal as well as the spread in pedestal has been determined
for each chip.

Fig.~7 shows (a) the distribution of pedestal minima and (b) scatter
plot of pedestal minima vs. pedestal spread for 5000 chips.  It is
seen that we can select chips of four categories having close ranges
of pedestal minima and pedestal spreads. The narrow width of the
distribution shows that the usable number of chips is a large fraction
of the total number of chips tested.

\begin{figure}[ht]
\setlength{\unitlength}{1mm}
\begin{picture}(70,70)
\put(0,0){
\epsfxsize=7cm
\epsfysize=7cm
\epsfbox{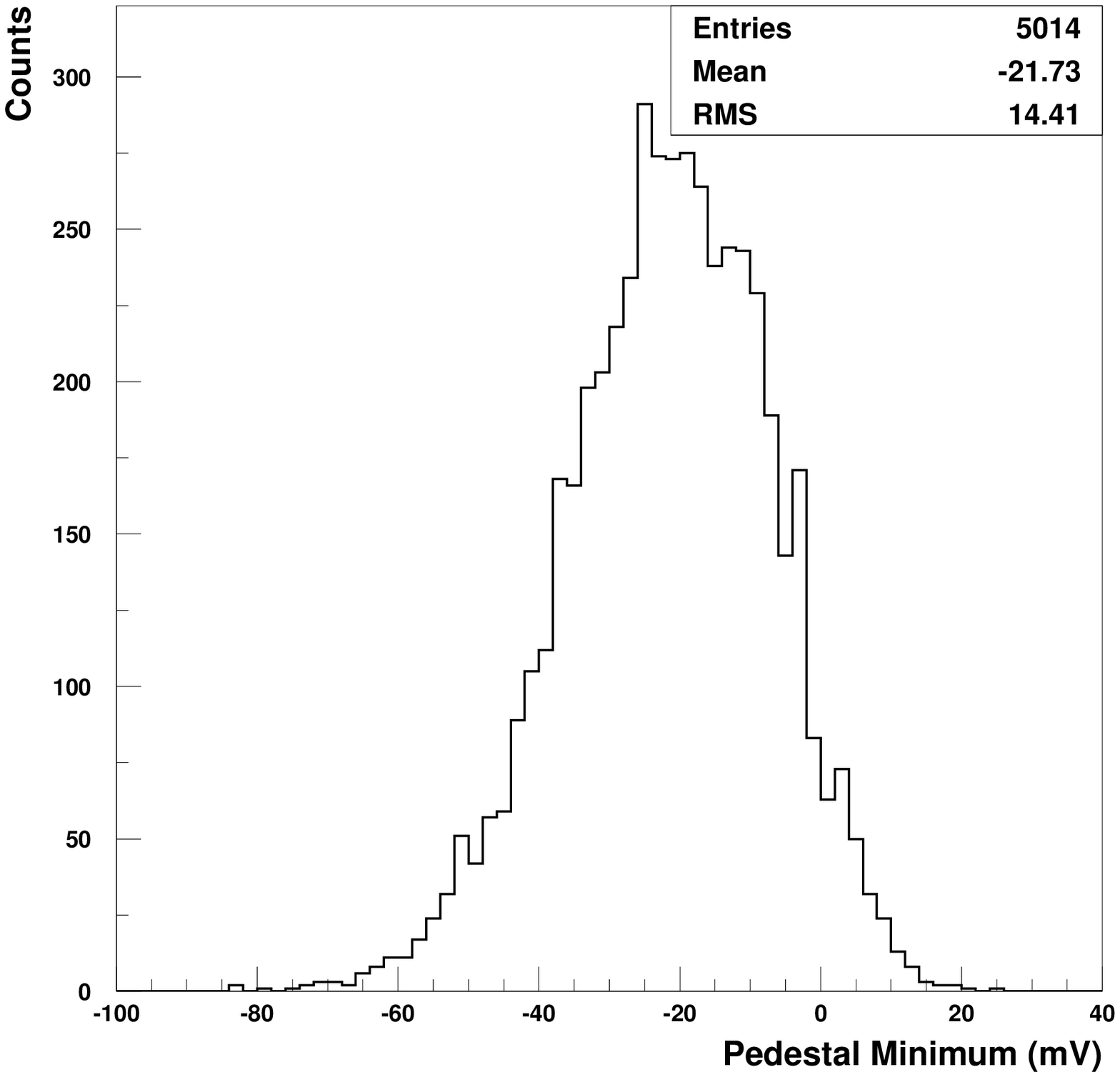}
}
\put(70,0){
\epsfxsize=7cm
\epsfysize=7cm
\epsfbox{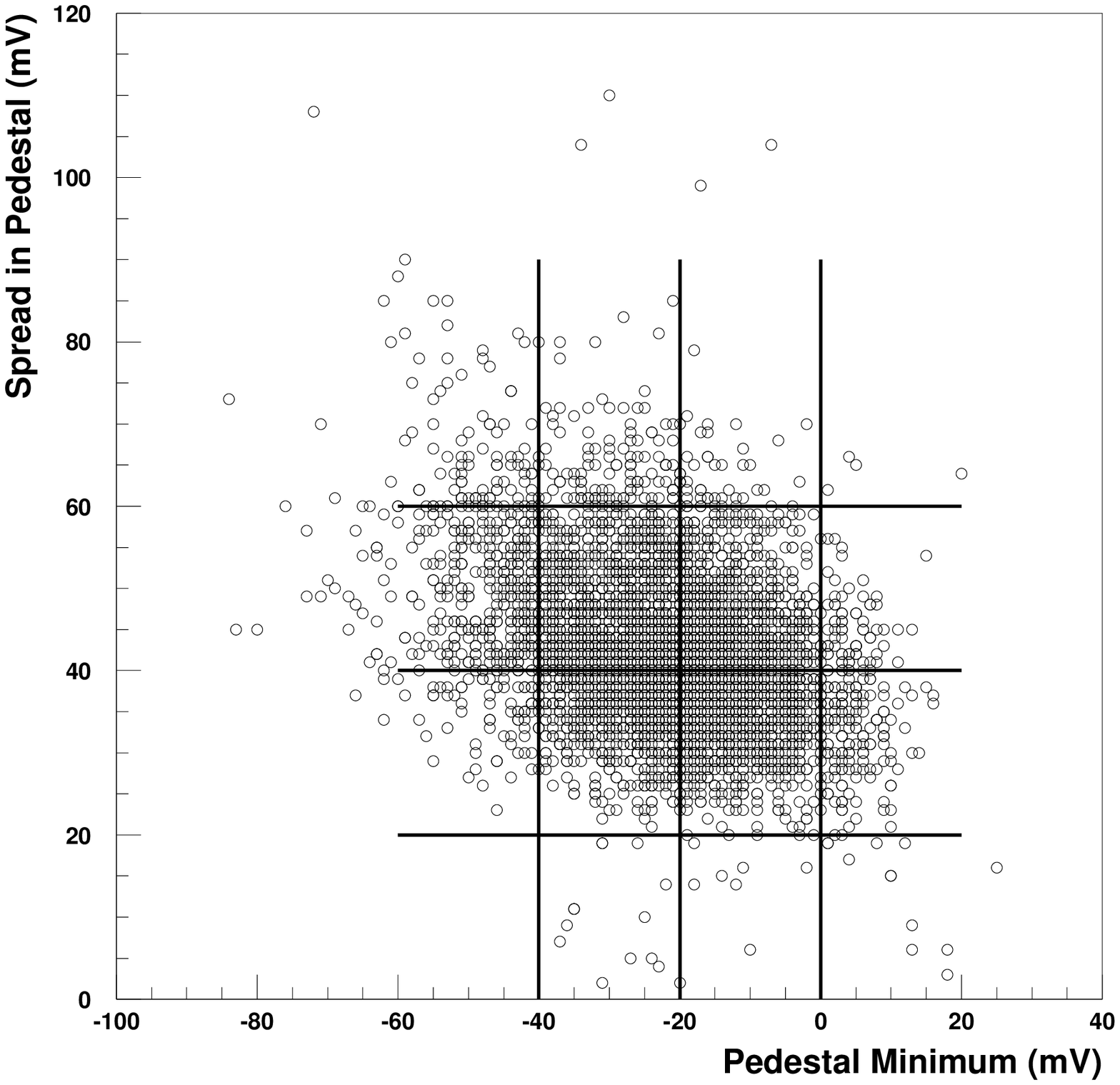}
}
\end{picture}
\caption{ (a) Pedestal minimum values ( in mV) for 5000
  chips, (b) pedestal minimum vs. pedestal spread for these chips.
  Lines are drawn to suggest the grouping of chips for a uniform
  chain.}
\label{mipcell}
\end{figure}

\section{Performance of the PMD}

Detailed tests have been performed with the prototype detector using
pion beam at the CERN PS for the study of the response of the minimum
ionizing particle (MIP) and electron beam with lead converter for
estimating the performance of the preshower characteristics of the
detector. Operating parameters, e.g., the composition of the gas
mixture and applied HV have been optimized by these tests. GEANT
simulation has been performed to study the effect of upstream material
on the physics performance of the PMD.  We discuss here some of the
results.

\vskip 0.5cm
\subsection{Response to charged particles}

Fig.~8(left) shows a typical MIP pulse height spectrum
with applied voltage of -1500~V. Fig.~8(right) shows the number of
cells hit by MIP, which is close to one.  This suggests that charged
hadron signal is essentially confined to one cell and satisfies one of
the main design criterion of the detector.


\begin{figure}[ht]
\setlength{\unitlength}{1mm}
\begin{picture}(70,70)
\put(0,0){
\epsfxsize=7cm
\epsfysize=7cm
\epsfbox{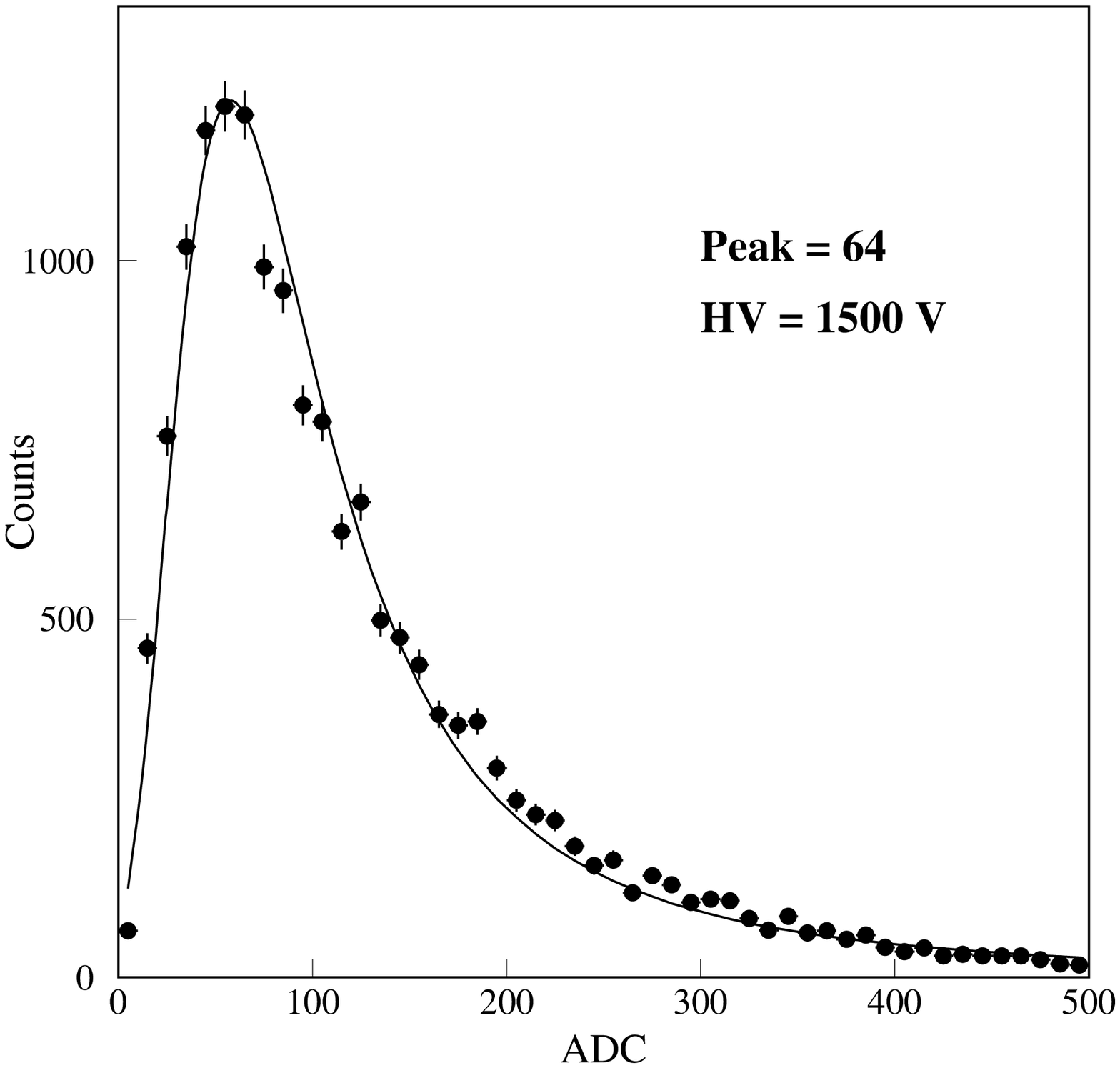}
}
\put(70,0){
\epsfxsize=7cm
\epsfysize=7cm
\epsfbox{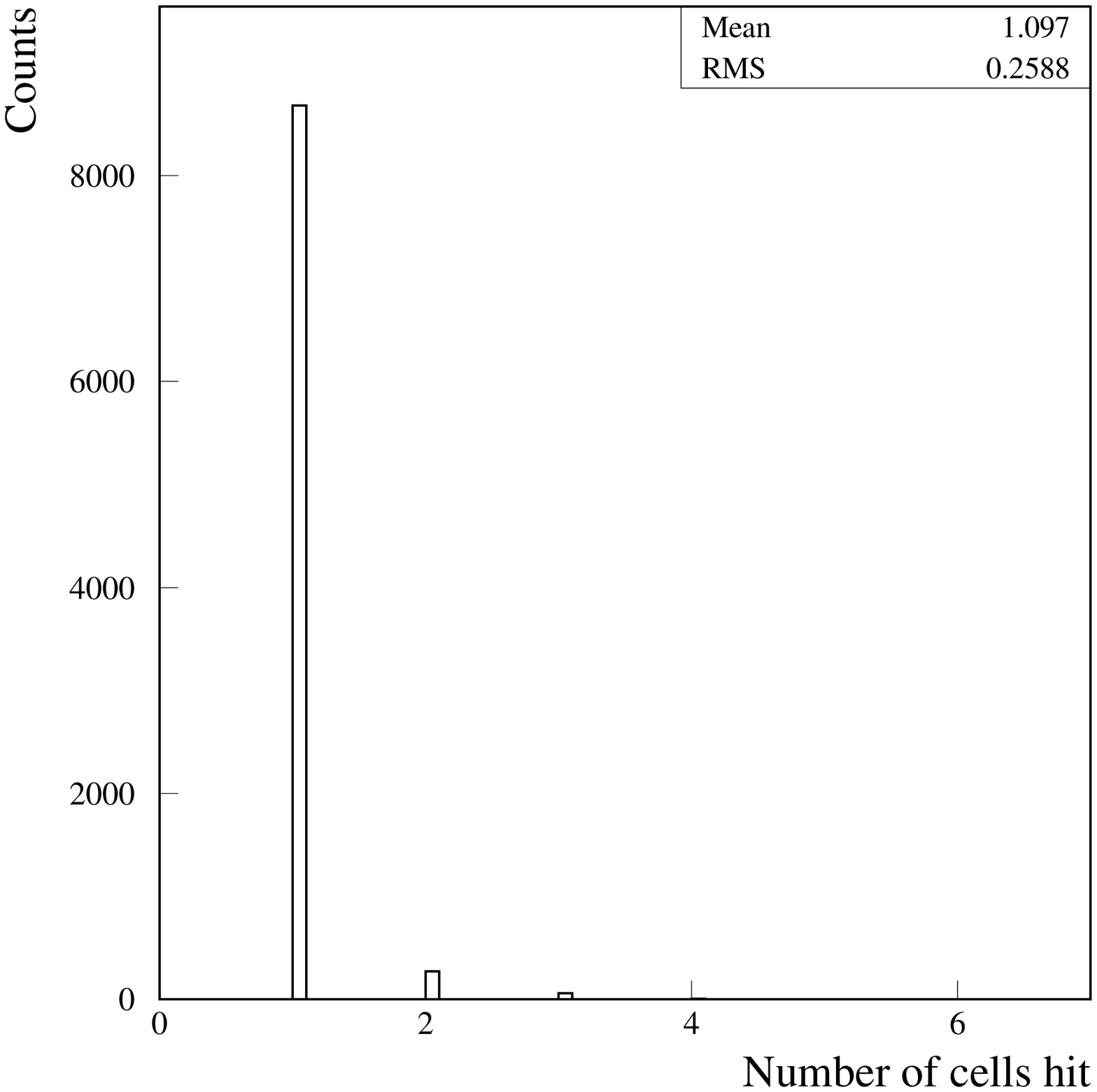}
}
\end{picture}

\caption{ (left) Typical MIP spectra for cells with -1500V , (right)
  Distribution of the number of cells hit by MIP, it is seen that MIP
  is contained mostly in one cell.}
\label{chiptest}
\end{figure}

The efficiency for charged particle detection and the gain of the cell
has been determined for a number of cells chosen randomly in the
prototype.  Fig.~9 (top) shows the histogram of the relative gains of
40 cells.  The relative gain is defined by the ratio of the mean pulse
height in a cell to the average value of the mean pulse heights of all
the 40 cells taken together. The overall gain of the prototype chamber
is found to be quite uniform, the distribution having a narrow width
of $\sigma \approx 6\%$.  Fig.~ 9(bottom) shows the efficiency for 40
cells. The average value of the efficiency is found to be 98\%. The
efficiency is also uniform over the cross-section of the hexagon
within a single cell, varying within a narrow range of 93--99\%.  the
lower value being at the edges of the cell.

\begin{figure}[ht]
\epsfxsize=14cm
\epsfbox{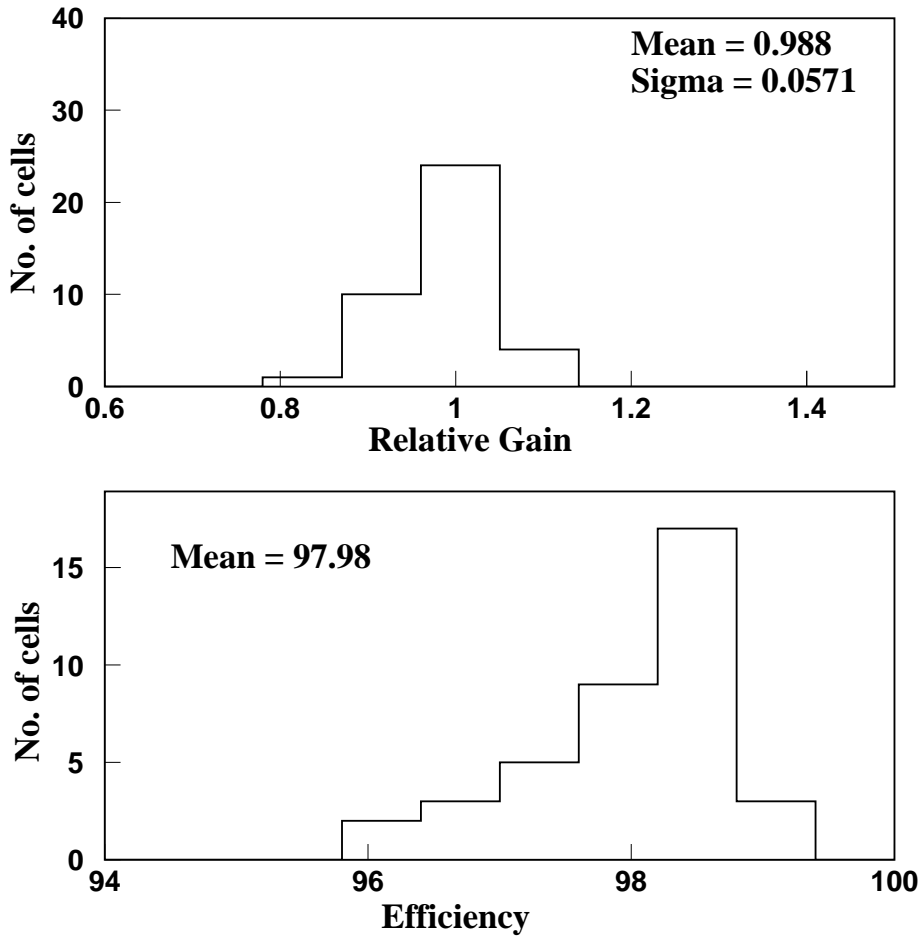}

\caption{Distributions of (top) gain and (bottom) efficiency for randomly selected 40 cells from the prototype chamber.}

\label{gaineff}
\end{figure}
\vskip 0.5cm
\subsection{Preshower Characteristics}

Preshower behaviour is characterised by (a) the transverse spread of
the shower, which is given by the size of the cluster of hit cells,
and (b) by the energy deposition expressed in terms of the cluster
signal (i.e., the total signal in all the hit cells, in ADC units).
These have been determined using 1--6 GeV electrons and a 3$X_0$ thick
lead converter kept in front of the prototype detector.
 
The typical preshower spread for 3~GeV electrons is shown in Fig.~10
(left). Cluster size obtained from test data is very close to the
values obtained from GEANT simulation, thereby suggesting that the
occupancy of the detector for a given multiplicity can be obtained
reliably with GEANT simulation.

\begin{figure}[ht]
\setlength{\unitlength}{1mm}
\begin{picture}(70,70)
\put(0,0){
\epsfxsize=7cm
\epsfysize=7cm
\epsfbox{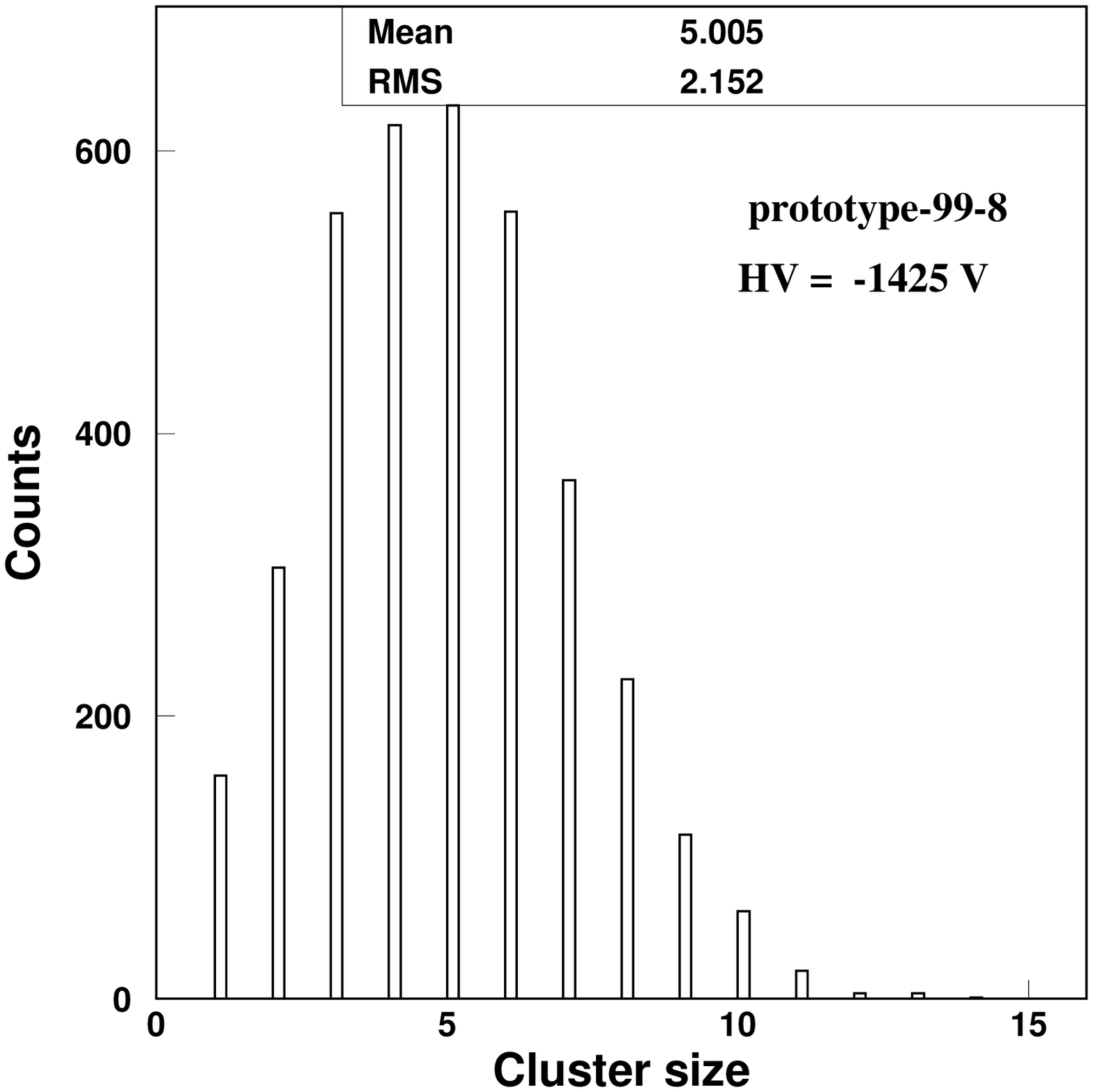}
}
\put(70,0){
\epsfxsize=7cm
\epsfysize=7cm
\epsfbox{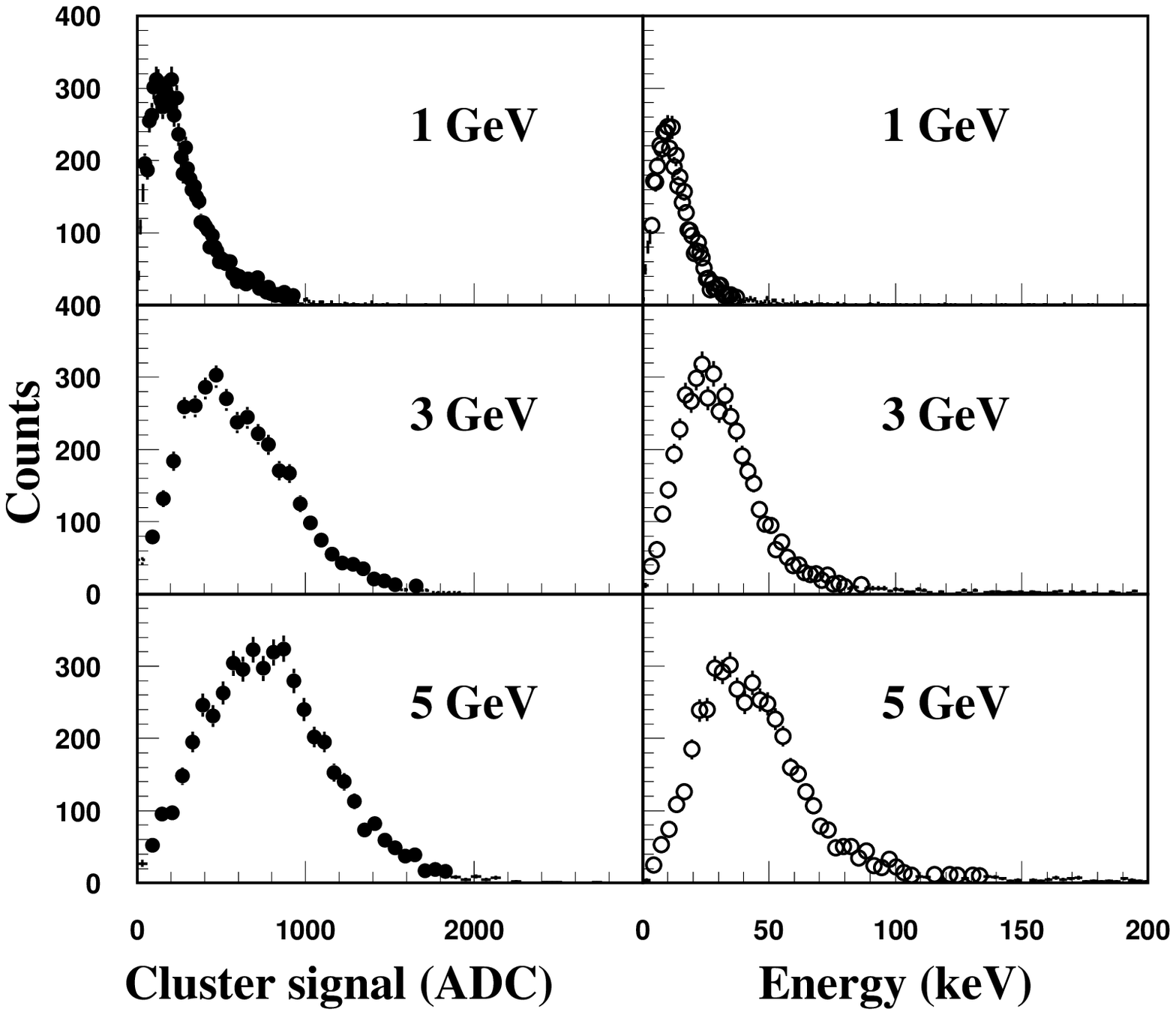}
}
\end{picture}
\caption{ (left) Typical cluster size for preshower expressed in terms
  of the cells affected by electron, (right panel, filled circle)
  energy deposition (in terms of cluster ADC) spectra for electrons
  with 3 different energies , (right panel, open circle) Simulated
  energy deposition (in keV) for electrons with corresponding
  energies. Width of simulated spectra is smaller compared to test
  data. }
\label{pscell}
\end{figure}


\vskip 0.5cm
\subsection{Simulation of the preshower response of the detector}

The energy deposition spectra for electrons at various energies as
obtained from test data and those obtained from the GEANT simulation
at corresponding energies are shown in Fig.~ 10(right panel).  Even
though the shapes look similar, the relative widths in the preshower
spectra are larger for test data compared to those in simulation.
This difference is due to fluctuations in gas ionization, signal
generation and transmission process in data, which are not accounted
for in simulation.  It was therefore necessary to estimate and
introduce this difference in widths with proper modeling.  We refer
this addition in spread to the simulated spectra as {\it readout
  width}.

Fig.~ 11 (left) shows the readout width for a range of energy
deposition values.  For this plot, data using 2~X$_0$ thick converter
have also been used.  From the given plot, we can deduce the readout
width for any given energy deposition obtained from GEANT simulation
and fold the values for detailed comparison with experimental data.

Fig.~ 11 (right) shows the mean energy deposition obtained from
simulation plotted against the mean ADC obtained for a particle of
given energy. The response of the detector and readout is seen to be
fairly linear in the range of energy studied, upto that expected from
10 GeV photons in the preshower part.

\begin{figure}[ht]
\setlength{\unitlength}{1mm}
\begin{picture}(70,70)
\put(0,0){
\epsfxsize=7cm
\epsfysize=7cm
\epsfbox{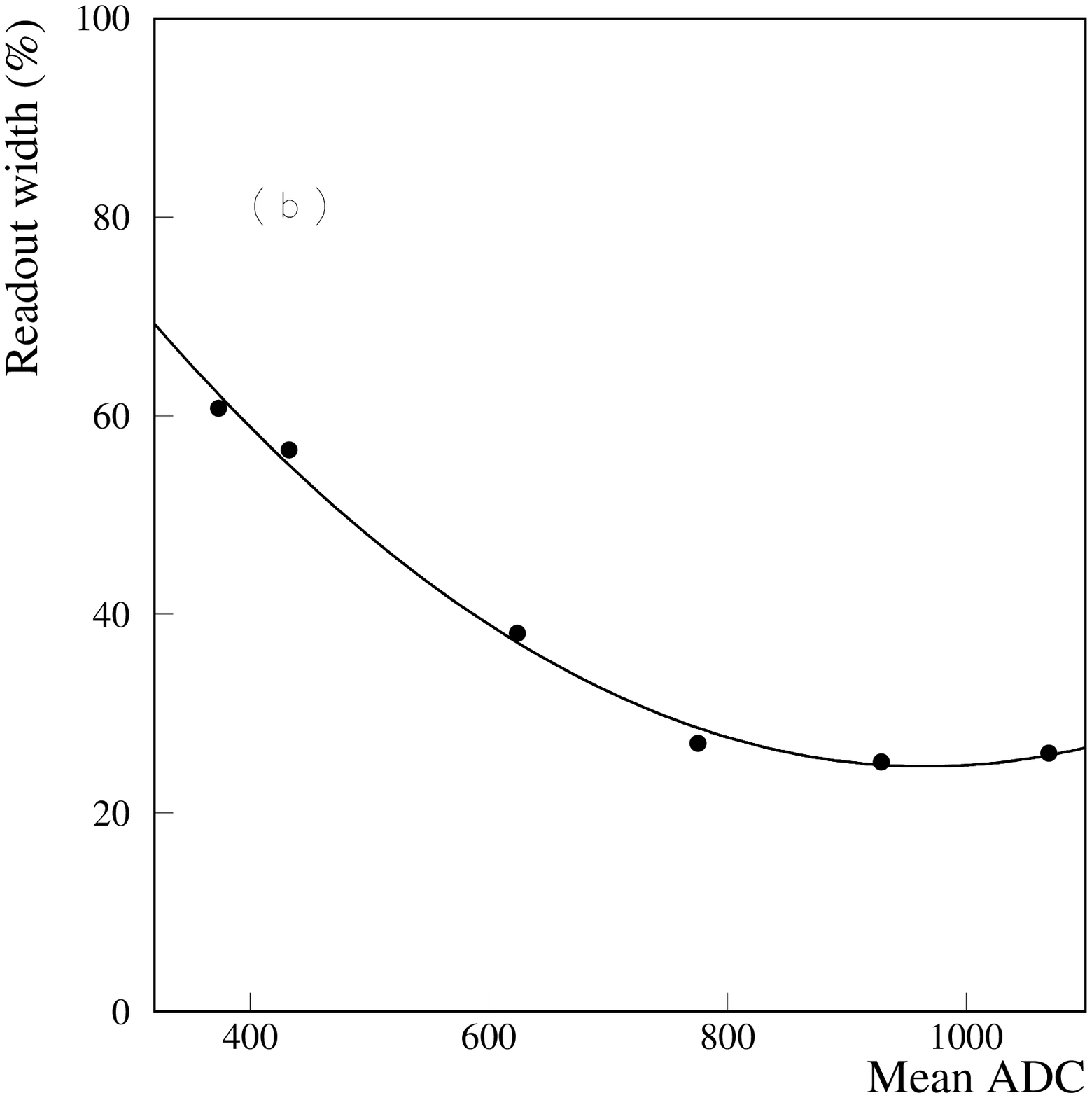}
}
\put(70,0){
\epsfxsize=7cm
\epsfysize=7cm
\epsfbox{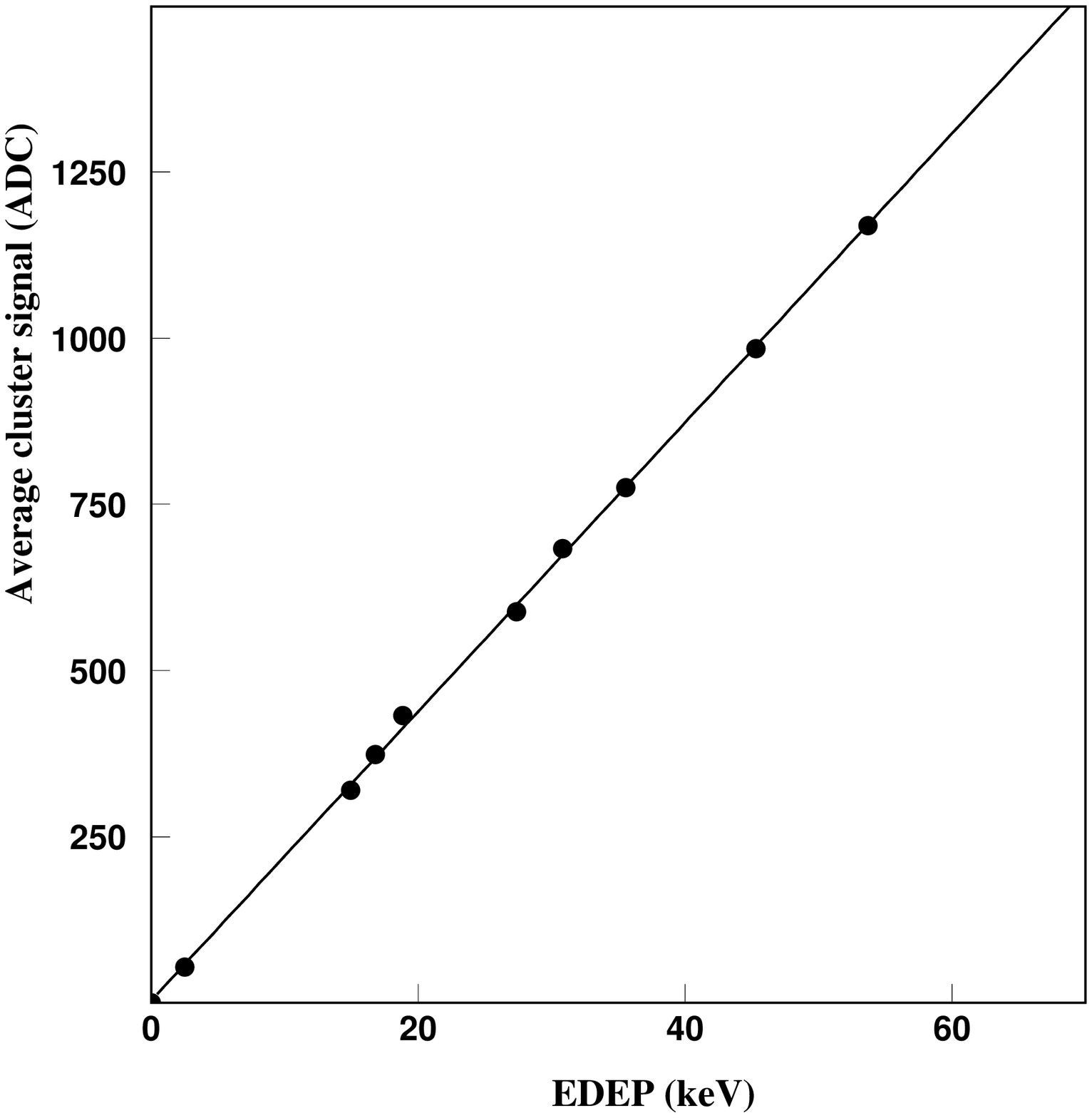}
}
\end{picture}
\caption{(left) Readout width ($\%$) shown for various energy 
  deposition expressed in terms of cluster ADC, (right) Calibration
  plot, showing the relation between the energy deposition obtained
  from simulation and the cluster ADC obtained from test data.}
\label{fig:readout}
\end{figure}

\subsection{Photon counting efficiency}

For counting of photons using the preshower PMD, clusters of hit
cells are determined and then a suitable algorithm is used for
discrimination of charged hadrons giving signal on the preshower
plane. Usually a small percentage of charged hadron signals pass
the discrimination test and mix with the photon sample, causing
the purity of the sample to decrease. Due to the presence of
upstream material from other STAR detectors, a part of photons 
may get converted and fall on the PMD as charged particles. In
addition high particle density causes overlap of clusters and
reduces the number of good photon clusters.

The physics performance of the preshower PMD is characterized by
two quantities : photon counting efficiency ($\epsilon_\gamma$) 
and purity ($f_p$) defined by the following 
relations \cite{wa98pmd}:

\begin{eqnarray} 
\epsilon_\gamma = N^{\gamma,{\rm th}} _{\rm cls} / N^\gamma _{\rm inc}~,
\end{eqnarray}
\vspace*{-0.6cm}
\begin{eqnarray} 
f_p = N^{\gamma,{\rm th}} _{\rm cls} / N_{\gamma{\rm -like}}~.
\end{eqnarray}

\noindent where     $N^\gamma _{\rm inc}$ is the   number of incident photons from the event
generator, $N^{\gamma,{\rm th}} _{\rm cls}$ is the number of photon
clusters above the hadron rejection threshold and $N_{\gamma{\rm
    -like}}$ is the total number of clusters above the hadron
rejection threshold.  (1$-f_p$) is the fractional contamination in the
$N_{\gamma{\rm -like}}$ sample.

Using GEANT simulation and an event generator, the photon counting
efficiency and purity for the PMD have been studied in the actual
environment of STAR experiment. 
 For the present case, both the efficiency and purity are
found to be around 60\%.


\section{Summary}

The preshower PMD, based on honeycomb proportional counter design,
will be an important addition to the STAR experiment.  The detector
consists of two identical planes of high granularity, one placed in
front of lead converter and acting as charged particle veto, and
another behind the converter acting as preshower detector. Each plane
has 12 supermodules, a supermodule being a gas-tight chamber
consisting of 4 to 9 unit modules of 576 cells arranged in a rhombus
matrix.

Prototypes have been tested to study charged particle detection and
preshower characteristics.  The performance satisfies the main design
criteria quite well. The cellular design is found to contain
$\delta-$electrons and minimize the spread of the signal to
neighboring cells. The charged particle signal is confined mostly to
one cell.  With the inclusion of extended cathode geometry, the
efficiency is quite uniform over the entire cell.

The preshower data show that the transverse shower size is in close
agreement with single particle GEANT simulations. Average pulse height
of the preshower follows a linear relation with energy deposition for
a wide range, upto that expected from 10 GeV photons in the preshower
part of the STAR PMD.

The fabrication procedure for unit modules and supermodules
incorporate quality control measures to guarantee the performance of
the whole detector as observed for the prototype.

The front-end electronics is based on the 16-channel analogue
multiplexed GASSIPLEX chips for signal processing and C-RAMS ADCs for
readout.  There are 48 readout chains, each having 1728 channels. FEE
boards containing four GASSIPLEX chips have been designed to allow
modularity.  Chips within a chain are selected, after functionality
and pedestal tests, to have close values of pedestal minima.

The PMD, covering the pseudorapidity region 2.3$\le \eta \le$3.5 with
full azimuthal acceptance and placed behind the FTPC, will be used to
study fluctuation, flow and chiral symmetry restoration. It will
measure the spatial distribution of photons with an efficiency of
about 60\% and a purity of about 60\%.

\vskip 1cm

{\bf Acknowledgements}

We wish to express our gratitude to the Department of Atomic Energy
and the Department of Science and Technology of the Government of
India for their support in this project. One of us (SD) acknowledges
the grant of research fellowship of the Council of Scientific and
Industrial Research in India. Permission by CERN to allow the use of
GASSIPLEX chips is gratefully acknowledged. The R$\&$D for the
detector was carried out as a joint program with ALICE PMD at CERN and
the test beam facilities of CERN have been used extensively. The
support of CERN staff for test beam is acknowledged. We are specially
thankful to J.C. Santiard for help in the development of front-end
electronics and readout scheme.

\end{document}